\documentclass[aps,prd,preprintnumbers,showpacs,superscriptaddress,nofootinbib,amsmath,amssymb,floats,floatfix,showkeys,notitlepage,longbibliography]{revtex4-1}
\usepackage{comment}
\usepackage{graphicx}
\usepackage{subfigure}
\usepackage{palatino}
\usepackage[commandnameprefix=always]{changes}
\usepackage{hyperref}
\hypersetup{colorlinks=true,linkcolor=red,urlcolor=red,citecolor=red}
\usepackage[toc,page]{appendix}
\usepackage[normalem]{ulem}

\usepackage{lipsum}
\usepackage{graphicx}
\usepackage{subfigure}
\usepackage{palatino}
\usepackage{float}
\usepackage{sans}
\usepackage{adjustbox}
\usepackage{latexsym}
\usepackage{amsmath}
\usepackage{amssymb}
\usepackage{amsfonts}
\usepackage{dcolumn}
\usepackage{bm}
\usepackage{tikz}
\usepackage{bigints}
\usepackage{array,tabularx,multirow,booktabs}
\usepackage[tracking=true]{microtype}
\SetTracking{}{500}
\SetTracking{encoding={*}, shape=sc}{40}
\allowdisplaybreaks
\usepackage{adjustbox}
\usepackage{latexsym}
\usepackage{amsmath}
\usepackage{amssymb}
\usepackage{amsfonts}
\usepackage{dcolumn}
\usepackage{bm}
\usepackage{tikz}
\usepackage{bigints}
\usepackage{array,tabularx,multirow,booktabs}
\usepackage[tracking=true]{microtype}
\usepackage{color}
\allowdisplaybreaks
\begin{document}

\title{Striving to find a bridge between noncommutative and generalized entropy parameters in harmonic oscillator dynamics}

\author{Jafar Sadeghi}
\email{pouriya@ipm.ir}
\affiliation{Department of Theoretical Physics, Faculty of Basic Sciences,
University of Mazandaran\\
P. O. Box 47416-95447, Babolsar, Iran}
\affiliation{School of Physics, Damghan University, P. O. Box 3671641167, Damghan, Iran}

\author{Mohammad Ali S. Afshar}
\email{Corresponding author-- m.a.s.afshar@gmail.com }
\affiliation{Department of Theoretical Physics, Faculty of Basic Sciences,
University of Mazandaran\\
P. O. Box 47416-95447, Babolsar, Iran}
\affiliation{School of Physics, Damghan University, P. O. Box 3671641167, Damghan, Iran}
\affiliation{Center for Theoretical Physics, Khazar University, 41 Mehseti Street, Baku, AZ1096, Azerbaijan}

\author{H.Z.Kalbasti}
\email{mahanparsao5000@gmai.com}
\affiliation{Department of Theoretical Physics, Faculty of Basic Sciences,
University of Mazandaran\\
P. O. Box 47416-95447, Babolsar, Iran}

\author{H.Razazchian}
\email{razazchianhirbood@gmai.com}
\affiliation{Department of Theoretical Physics, Faculty of Basic Sciences,
University of Mazandaran\\
P. O. Box 47416-95447, Babolsar, Iran}

\begin{abstract}
In this study,  we use the harmonic oscillator’s energy spectrum to connect its dynamical behavior to the formalism of  statistical mechanics through entropy. We perform a thermal analysis of the entropy generated by this spectrum in both commutative and noncommutative two-dimensional configuration spaces, employing the Tsallis and R\'enyi entropy measures. \\ 
The primary objective is to examine the possibility of an intrinsic link and mutual influence between the noncommutative parameter($\theta$) and the parameters arising from generalized entropies. By analyzing both Tsallis and R\'enyi frameworks, we investigate whether a fundamental relationship exists between $\theta$ and the entropy index $q$. Establishing such a connection could provide new insights into the interplay between spacetime geometry and generalized statistical mechanics.
\end{abstract}
\date{\today}
\maketitle
\section{Introduction}
At the dawn of quantum theory, pioneers such as Heisenberg, Born, and Jordan formulated matrix mechanics, in which observables like position \(x\) and momentum \(p\) obey the non-commutation relation  
\begin{equation}
[x, p] = i\hbar
\end{equation} 
instead of commuting as in classical physics \([x, p] = 0\). This fundamental departure underlies the uncertainty principle and distinguishes quantum dynamics from its classical counterpart. The emergence of this commutation rule laid the theoretical cornerstone of quantum mechanics and gave rise to Heisenberg’s uncertainty principle \cite{1}.\\
Building on matrix mechanics, these founders extended non-commutativity to fields such as the electromagnetic field. In quantum field theory (QFT), a scalar field \(\phi(x)\) and its conjugate momentum \(\pi(y)\) satisfy  
\begin{equation}
[\phi(x), \pi(y)] = i\,\delta(x - y)\,,
\end{equation}  
thereby generalizing the position–momentum commutator to spacetime \cite{2}. This step established QFT as the natural framework for describing particle creation and annihilation while preserving quantum uncertainty at every spacetime point.\\
Decades later, In the 1980s, Alain Connes introduced noncommutative geometry (NCG), a mathematical framework where spacetime coordinates \(x^\mu\) themselves fail to commute \cite{3}:  
\begin{equation}
[x^\mu, x^\nu] = i\,\theta^{\mu\nu}\,,
\end{equation}  
with \(\theta^{\mu\nu}\) an antisymmetric matrix encoding a fundamental “fuzziness.”\\ 
This construction mirrors the phase-space uncertainty in quantum mechanics and introduces an intrinsic minimal length scale of order $ \sqrt{|\theta^{\mu\nu}|} $, suggesting a natural ultraviolet (UV) cutoff in field theories defined on such spaces \cite{4}.\\
Inspired by Connes’ work, physicists formulated noncommutative field theories (NCFTs), where the standard pointwise product of fields is replaced by the Moyal–Weyl star product:
$$
(f \star g)(x) = f(x)\, e^{\frac{i}{2}\,\theta^{\mu\nu} \overleftarrow{\partial_\mu} \overrightarrow{\partial_\nu}}\, g(x),
$$
which encodes the non-locality induced by spacetime non-commutativity \cite{5}. This modification alters the Feynman rules of the theory, leading to a phenomenon known as UV/IR mixing—where ultraviolet divergences are intertwined with infrared effects—but also has the potential to regularize high-energy behavior. Notably, the noncommutative scale $ \sqrt{\theta} $ acts as a geometric regulator, offering a physically motivated alternative to ad hoc momentum cutoffs used in conventional renormalization procedures \cite{6}.\\
This philosophy extends beyond field theory into string theory and quantum gravity. When a D-brane is immersed in a strong background \(B\)-field, the endpoints of open strings on the brane obey  
\begin{equation}
[x^i, x^j] = i\,\theta^{ij}\,,
\end{equation}  
where $ \theta^{ij} \propto (B + F)^{-1} $, with $ F $ being the field strength of the gauge field on the brane. This result is not an imposed regulator but a natural consequence of open-string dynamics and the Seiberg–Witten duality, which relates noncommutative and commutative field theories. Depending on the physical regime, one may equivalently describe the worldvolume theory in either picture \cite{7}. Noncommutative spacetime structures have found applications across various scales, from inflation \cite{7.1}, cosmology \cite{7.2,7.3} and black holes \cite{7.4,7.5,7.6,7.7,7.8,7.9} to subatomic systems \cite{7.11,7.12,7.13}.\\
The finite-temperature thermodynamics of the noncommutative harmonic oscillator has been previously studied using the standard Boltzmann-Gibbs framework via path integral and Hamiltonian methods \cite{Jahan2012,Lin2018}. These works established the modification of the partition function and free energy due to noncommutativity. In the present study, we extend this analysis by employing generalized entropy measures (Tsallis and R\'enyi) to explore possible parametric connections between the noncommutative parameter $\theta$ and the entropic index $q$.\\\\
From the beginning of the quantum mechanics, a widespread method for analyzing quantum systems involves mapping their Hamiltonians onto that of a harmonic oscillator. Operator techniques and path-integral formulations are then used to solve for the spectrum. In a noncommutative setting:\\
- The oscillator’s energy levels shift: \(E_n \to E_n(\theta)\).\\  
- The partition function becomes  
 \begin{equation}
  Z(\theta) = \sum_{n=0}^{\infty} e^{-\beta E_n(\theta)}\,.
 \end{equation} 
- The entropy generalizes to  
 \begin{equation}
  S(\theta) = -k_B \sum_{n=0}^{\infty} P_n \,\ln P_n,\quad
  P_n = \frac{e^{-\beta E_n(\theta)}}{Z(\theta)},
 \end{equation}
where $\beta=1/k_{B} T$ , $k_{B}$ is Boltzmann coefficient, $T$ temperature and  $P_{n}$ is the probability of the system being in the n-th microstate \cite{8}.\\
Since the energy spectrum links dynamics to statistical mechanics via entropy, noncommutative modifications open a direct pathway into thermodynamics and statistical physics.
Historically, Boltzmann–Gibbs entropy underpinned classical statistical mechanics, on which most of the early analyses and explanations were based. However, in systems with long-range interactions, memory effects, or fractal phase spaces, this entropy  is no longer sufficient  to capture the full thermodynamic behavior. Over time, these weaknesses led to the introduction of the alternative generalized entropies —Tsallis, R\'enyi, Shannon, Sharma–Mittal, and others— which have been gradually proposed. \cite{9,10,11}. Investigating how spacetime non-commutativity influences these generalized entropies constitutes a novel research frontier. Specifically, exploring whether a substantive relationship exists between the non-commutativity parameter $ \theta $ and the Tsallis entropic index $ q $ could reveal deep connections between geometric spacetime structure and statistical mechanics. Such a connection  could provide insights into quantum statistical systems in noncommutative settings and may be relevant for understanding  thermodynamic aspects of models with minimal length scales.\\
Motivated by these considerations, in this work, we analyze a harmonic oscillator subject to noncommutative geometry and replace the conventional Boltzmann–Gibbs entropy with Tsallis entropy, exploring whether a substantive relationship emerges between the non-commutativity parameter $\theta$ and the Tsallis index $q$ and assessing how this interplay bridges established results with novel theoretical insights.
\section{Simple Harmonic Oscillator}
Since our analysis of the statistical form is going to be based on the energetic behavior of the two-dimensional isotropic harmonic oscillator, it is better to begin the discussion by introducing the equations of this oscillator in two separate cases.\\\\
\textbf{1) General Form}:\\\\
The usual Hamiltonian structure for the two-dimensional isotropic harmonic oscillator is given by: 
\begin{equation}
\hat{H} = \frac{\hat{p}_x^2 + \hat{p}_y^2}{2m} + \frac{1}{2} m \omega^2 (\hat{x}^2 + \hat{y}^2).
\label{1}
\end{equation}
Based on the traditional standard method the standard ladder (annihilation and creation) operators can be written as follows:

\begin{align}
a_x &= \sqrt{\frac{m\omega}{2\hbar}} \left( \hat{x} + \frac{i}{m\omega} \hat{p}_x \right), \quad
a_x^\dagger = \sqrt{\frac{m\omega}{2\hbar}} \left( \hat{x} - \frac{i}{m\omega} \hat{p}_x \right), \\
a_y &= \sqrt{\frac{m\omega}{2\hbar}} \left( \hat{y} + \frac{i}{m\omega} \hat{p}_y \right), \quad
a_y^\dagger = \sqrt{\frac{m\omega}{2\hbar}} \left( \hat{y} - \frac{i}{m\omega} \hat{p}_y \right).
\label{2}
\end{align}

Using these operators, the Hamiltonian Eq. (\ref{1}) takes the form:

\begin{equation}
\hat{H} = \hbar \omega (a_x^\dagger a_x + a_y^\dagger a_y + 1),
\label{3}
\end{equation}
which yields the energy eigenvalues:
\begin{equation}
E_{n_x, n_y} = \hbar \omega (n_x + n_y + 1).
\label{4}
\end{equation}
In order to simplify the statistical calculations and focus on a symmetric configuration, we consider the special case where \( n_x = n_y = n \). Under this assumption, the energy spectrum reduces to:
\begin{equation}
E_n = \hbar \omega (2n + 1).\\
\label{5}
\end{equation}

\textbf{2) SIMPLE HARMONIC OSCILLATOR IN NONCOMMUTATIVE SPACE}:\\\\
In the noncommutative space, where the coordinate operators satisfy the commutation relation:
\begin{equation}
[\hat{x}, \hat{y}] = i\theta,
\label{6}
\end{equation}
where $\theta$ being the noncommutative parameter, to better analyze the system, one can perform the following coordinate transformation \cite{12}:
\begin{equation}
\hat{x} = x - \frac{\theta}{2\hbar} p_y, \quad \hat{y} = y + \frac{\theta}{2\hbar} p_x.
\label{7}
\end{equation}
Substituting these into the standard Hamiltonian of the harmonic oscillator Eq. (\ref{1}), we obtain:
\begin{equation}
\hat{H} = \left( \frac{1}{2m} + \frac{m\omega^2\theta^2}{8\hbar^2} \right)(p_x^2 + p_y^2) + \frac{1}{2}m\omega^2(x^2 + y^2) - \frac{m\omega^2\theta}{2\hbar}(x p_y - y p_x).
\label{8}
\end{equation}
To simplify this Hamiltonian, we define an effective mass $M$ and an effective frequency $\Omega$ as follows:
 \[
\left\{
\begin{aligned}
&\frac{1}{2M}= \frac{1}{2m} + \frac{m\omega^2\theta^2}{8\hbar^2}\nonumber,\\
& M\Omega^2 = m\omega^2,\\
&\Omega = \omega \left(1+\frac{m^2\omega^2}{4\hbar^2}\theta^2\right)^\frac{1}{2}\nonumber. \\
 \end{aligned}
\right.
\] 

Using the above relations, the Hamiltonian Eq. (\ref{8}) then becomes:
\begin{equation}
\hat{H} = \frac{1}{2M}(p_x^2 + p_y^2) + \frac{1}{2}M\Omega^2(x^2 + y^2) - \frac{M\Omega^2\theta}{2\hbar}(x p_y - y p_x).
\label{10}
\end{equation}
one can now introduce the modified creation and annihilation operators as \cite{12}:
\begin{align}
a_x &= \sqrt{\frac{M\Omega}{2\hbar}}\left(x + \frac{i}{M\Omega}p_x\right), &
a_x^\dagger &= \sqrt{\frac{M\Omega}{2\hbar}}\left(x - \frac{i}{M\Omega}p_x\right), \\
a_y &= \sqrt{\frac{M\Omega}{2\hbar}}\left(y + \frac{i}{M\Omega}p_y\right), &
a_y^\dagger &= \sqrt{\frac{M\Omega}{2\hbar}}\left(y - \frac{i}{M\Omega}p_y\right).
\label{11}
\end{align}
Expressed in terms of these operators, the Hamiltonian takes the form:
\begin{equation}
\hat{H} = \hbar\Omega(a_x^\dagger a_x + a_y^\dagger a_y + 1) - \frac{M\Omega^2\theta}{2i}(a_x^\dagger a_y - a_y^\dagger a_x).
\label{12}
\end{equation}
Due to the Schwinger representation for the angular momentum \cite{12}:
\begin{equation}
J_1 = \frac{1}{2}(a_x^\dagger a_y + a_y^\dagger a_x),\\
J_2 = \frac{1}{2i}(a_x^\dagger a_y - a_y^\dagger a_x),\\
J_3 = \frac{1}{2}(a_x^\dagger a_x - a_y^\dagger a_y),\\
\label{13.1}
\end{equation}
the Hamiltonian can be rewritten in terms of the angular momentum-like operator in the compact form as:
\begin{equation}
\hat{H} = \hbar\Omega(N_x + N_y + 1) - M\Omega^2\theta J_2.
\label{13}
\end{equation}

In a suitable basis that diagonalizes both $N'_x + N'_y$ and $N'_x - N'_y$, the Hamiltonian simplifies to:
\begin{equation}
\hat{H} = \hbar\Omega(N'_x + N'_y + 1) - \frac{M\Omega^2\theta}{2}(N'_x - N'_y).
\label{14}
\end{equation}

Therefore, the energy eigenvalues of the oscillator in noncommutative space are given by:
\begin{equation}
E_{n_x, n_y} = \hbar\Omega(n_x + n_y + 1) - \frac{M\Omega^2\theta}{2}(n_x - n_y).
\label{15}
\end{equation}
This result shows that the degeneracy of the energy levels is lifted due to non-commutativity, and the spectrum explicitly depends on the difference $n_x - n_y$. In the limit $\theta \to 0$, the correction term vanishes and we recover the usual energy spectrum of the isotropic two-dimensional harmonic oscillator in commutative space.
\section{Statistical and Thermodynamic Properties }
\subsection{Simple harmonic oscillator and Tsallis entropy}
To investigate the thermodynamic properties, we adopt the Tsallis entropy ($S_{TE}$), defined as \cite{9,11}:
\begin{equation}
S_{TE} = \frac{k}{q - 1} \left(1 - \sum_i P_i^q \right),
\label{16}
\end{equation}
where \( P_i \) denotes the generalized probability distribution and 
$k$ is a positive constant. The corresponding generalized partition function under Tsallis statistics is:
\begin{equation}
Z_q = \sum_n \left[1 - \beta E_n (1 - q) \right]^{\frac{1}{1 - q}}.
\label{17}
\end{equation}

Substituting the simplified energy levels Eq. (\ref{5}), into the  above partition function we have:
\begin{equation}
Z_q = \sum_n \left[1 - \beta \hbar \omega (2n + 1)(1 - q) \right]^{\frac{1}{1 - q}}.
\label{18}
\end{equation}
In the high-temperature or small-\( \beta \) limit, this sum can be approximated by an integral, yielding the closed-form:
\begin{equation}
Z_q = \frac{1}{2 \beta \hbar \omega} \frac{\left( 1 - \beta \hbar \omega (q - 1) \right)^{\frac{1}{q - 1} + 1}}{ \left( \frac{1}{q - 1} + 1 \right)}.
\label{19}
\end{equation}
The normalized probability for a given energy level is then expressed as:
\begin{equation}
P_i = \frac{ \left( 1 - \beta E_i (1 - q) \right)^{\frac{1}{1 - q}} }{Z_q}.
\label{20}
\end{equation}
By substituting this into the entropy definition, we obtain the Tsallis entropy in terms of temperature and deformation parameter \( q \) as:
\begin{equation}
S_{TE} = \frac{k}{q - 1} \left[ 1 - \left( \frac{1 - \beta \hbar \omega (q - 1)}{2 \beta \hbar \omega (q - 1)} \right)^{1 - q}
\frac{\left(\frac{1}{q - 1} + 1 \right)^q} {1+\frac{q}{q - 1}} \right].
\label{21}
\end{equation}
To provide a comparison, we also include the standard Boltzmann-Gibbs (BG) result, which is recovered in the limit \(q \to 1 \). For the same energy spectrum:
\begin{equation}
E_n = \hbar \omega (2n + 1),
\label{22}
\end{equation}
and for the canonical partition function we have:
\begin{equation}
Z_{\text{BG}} = \sum_n e^{-\beta E_n} = \frac{e^{-\beta \hbar \omega}}{1 - e^{-2\beta \hbar \omega}}.
\label{23}
\end{equation}
Using the above relations, the corresponding BG entropy is given by:
\begin{equation}
S_{\text{BG}} = k \left[ \frac{2\beta \hbar \omega}{e^{2\beta \hbar \omega} - 1} - \ln \left( 1 - e^{-2\beta \hbar \omega} \right) \right].
\label{24}
\end{equation}
\begin{figure}[H]
 \begin{center}
 \subfigure[]{
 \includegraphics[height=5.5cm,width=7cm]{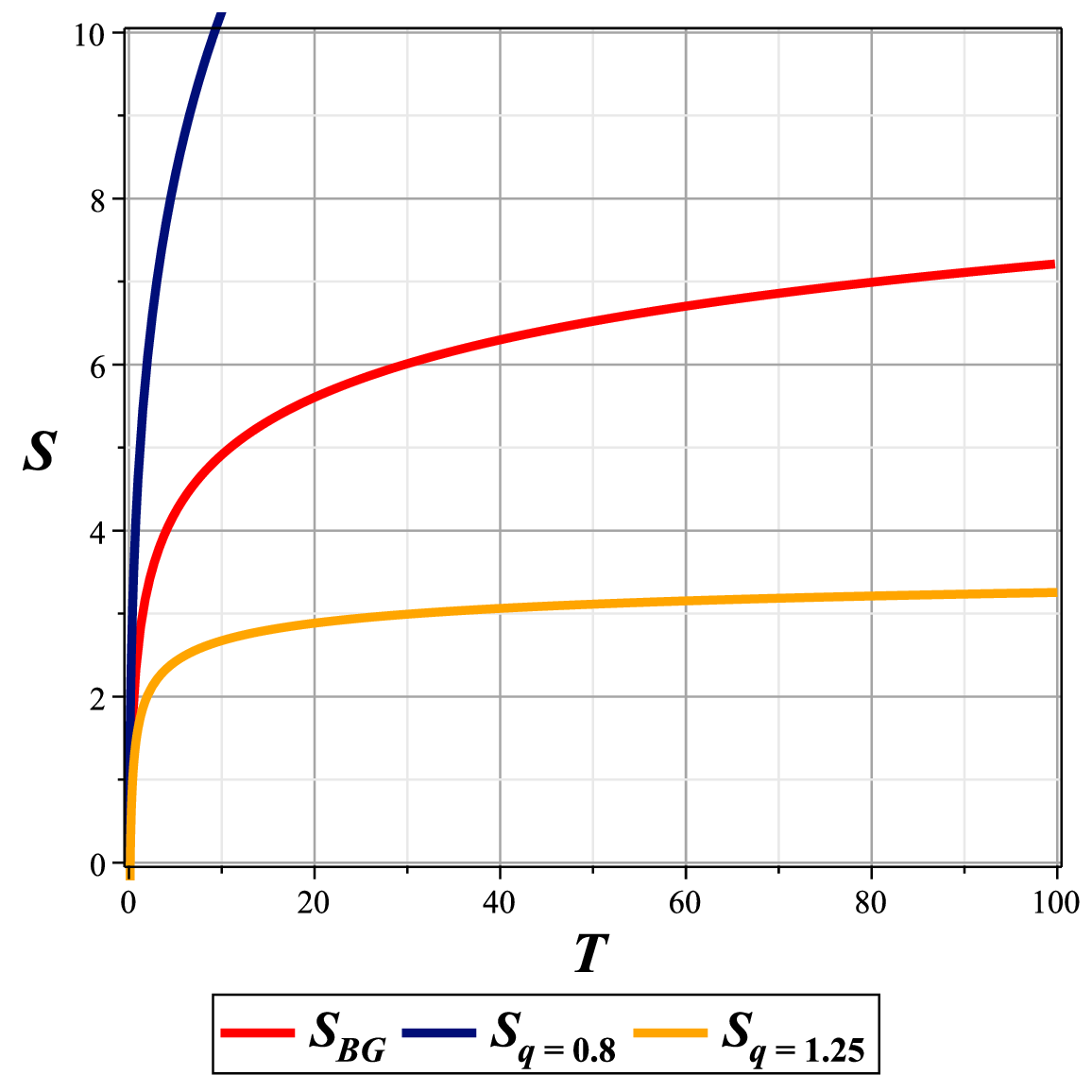}
 \label{1a}}
 \subfigure[]{
 \includegraphics[height=5.5cm,width=7cm]{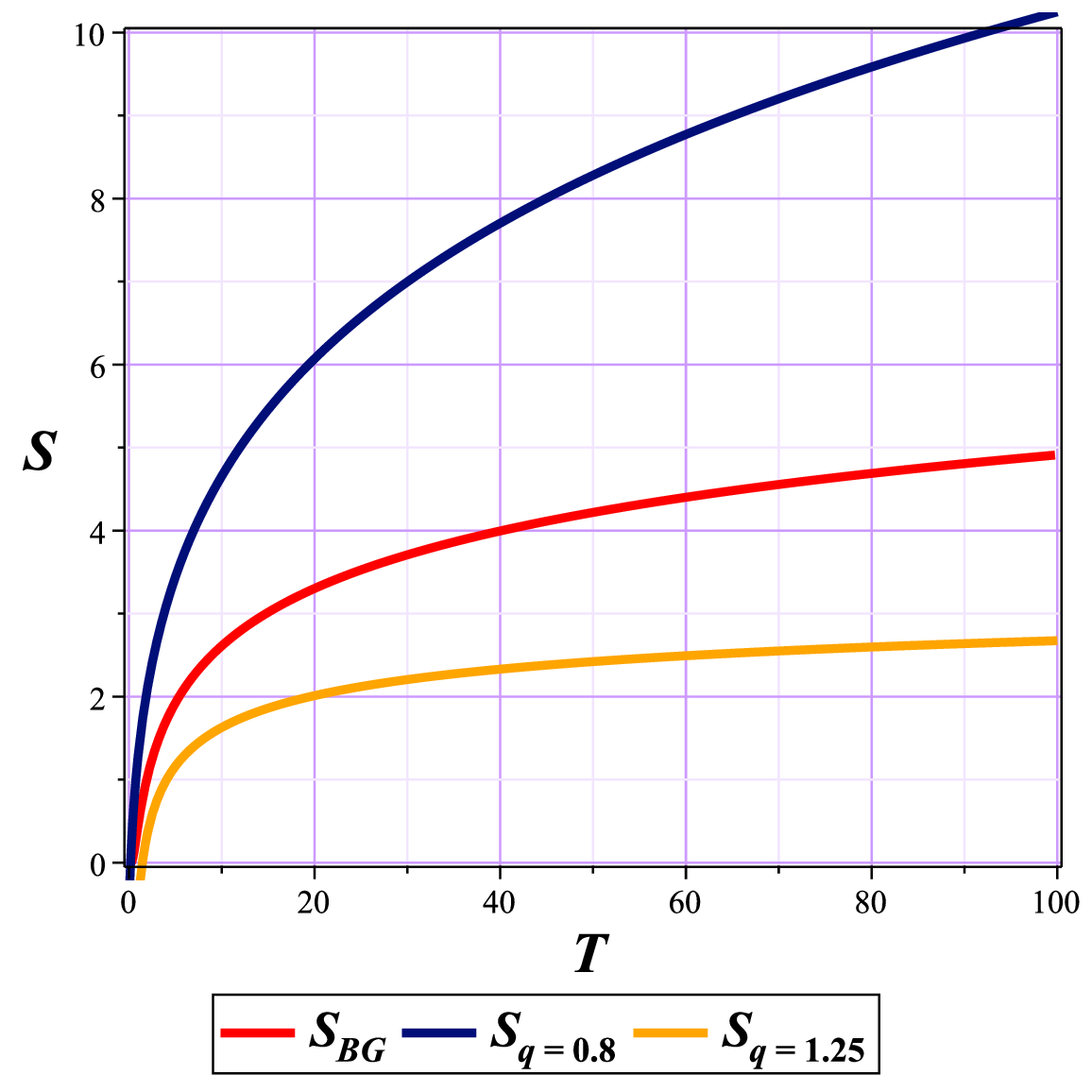}
 \label{1b}}
 \subfigure[]{
 \includegraphics[height=5.5cm,width=7cm]{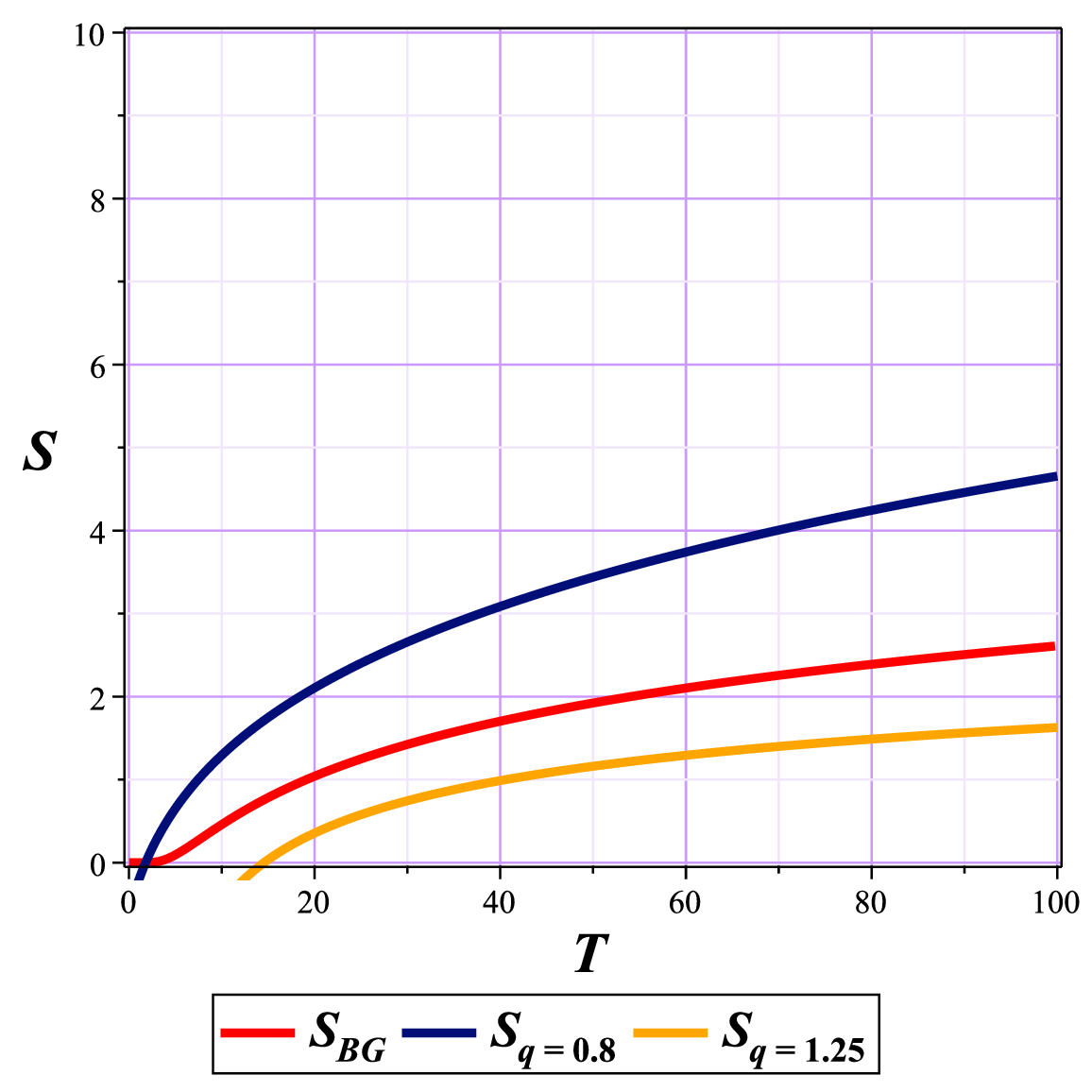}
 \label{1c}}
 
   \caption{\small{Temperature dependence of entropy for $\hbar=k=1$ and different values of $q$ and $\omega$. Fig (1a) $\omega=0.1$,  Fig(1b) $\omega=1$,  Fig(1c) $\omega=10$.  }}
 \label{(1)}
\end{center}
\end{figure}
One of the most notable advantages of employing generalized entropy forms lies in their inherent parametric freedom. These entropies introduce one or more additional parameters into the thermodynamic framework, functioning as tunable quantities under the experimentalist’s control. This feature becomes particularly valuable when the empirical behavior of a system deviates from predictions based on Boltzmann–Gibbs statistics. In such cases, conventional statistical model offer limited recourse, often forcing reliance on unrealistic mathematical approximations or simplifications that may or may not yield meaningful results.\\ In contrast, the inclusion of controllable parameters—derived through logical generalizations of the Boltzmann framework—enables one to accommodate system-specific constraints and calibrate the model within an acceptable range to obtain physically consistent interpretations. This parametric flexibility seems to be an important motivation behind the development of generalizedentropy formulations.\\
Intuitively, these entropy models are expected to exhibit systematic deviations from the baseline Boltzmann behavior as the parametric values vary, reflecting their broader descriptive capacity. This trend is clearly illustrated in Fig. (\ref{(1)}), where the Tsallis entropy, for values of \( q < 1 \) and \( q > 1 \), displays pronounced deviations from the Boltzmann–Gibbs entropy (represented by the red line). It is essential, however, that such entropy models revert to the standard form in the appropriate limit. For Tsallis entropy, this occurs as \( q \rightarrow 1 \), whereby the entropy converges to its Boltzmann–Gibbs counterpart. As Fig. (\ref{(1)}) demonstrates, the entropy curves for \( q = 0.8 \) and \( q = 1.25 \) symmetrically approach the classical line at \( q = 1 \), confirming this limiting behavior.\\
Furthermore, since the angular frequency \( \omega \) explicitly appears in the entropy of a quantum harmonic oscillator, we analyzed the impact of varying \( q \) across three distinct frequency regimes to better understand its influence. The results, shown in Fig. (\ref{(1)}), reveal that deviations are most prominent at low frequencies and gradually diminish as the frequency increases. Here, the term "deviations" specifically refers to the difference between the Tsallis entropy curves and the corresponding Boltzmann–Gibbs entropy curve. Moreover, the relative deviation decreases as the oscillator frequency increases. This behavior can be understood from Eq. (\ref{21}), where the entropy corrections induced by the nonextensive parameter \( q \) become progressively less significant at larger values of \( \omega \), causing the generalized entropy curves to approach the Boltzmann–Gibbs result.\\
\subsection{Simple harmonic oscillator and Rényi Entropy}
It is well established that Tsallis entropy, a generalization of Boltzmann entropy, does not inherit the property of additivity. In fact, its formulation was motivated by a fundamental question: What should be done when the system under study lacks additive entropy, and the total entropy changes upon combining subsystems?\\
The formation of studies based on Tsallis entropy led to further investigation into whether, under optimal conditions and through the application of Tsallis entropy—particularly in the presence of the entropic index$q$, it might be possible to restore additivity to the entropy formulation.\\
R\'enyi  entropy, which can be viewed as the logarithmic counterpart of Tsallis entropy, appears to offer a promising resolution to this issue. This is the distinctive property of R\'enyi entropy since, in addition to being armed with an additional parameter ($q$), it also maintains the additivity property.\\
The general form of R\'enyi entropy is given by \cite{10,11}:\\
\begin{equation*}\label{(0)}
S_q^{\text{Rényi}} = \frac{1}{1 - q} \ln \left( \sum_{i} p_i^q \right).
\end{equation*}
Now, on the one hand, since we have just calculated the Tsallis relations and compared them with the Boltzmann form, and on the other hand, since the following direct relation, namely:
\begin{equation}\label{(25)}
S_{\mathit{RE}}=\frac{\ln \! \left(1+S_{\mathit{TS}} \left(1-q \right)\right)}{1-q},
\end{equation}
can easily transfer us to the R\'enyi entropy, we can examine and observe the degree of deviation of these three entropy models from each other well.
With respect to Eq. (\ref{21}) for R\'enyi entropy we have:
\begin{equation}\label{26}
S_{\mathit{RE}}=-\frac{\ln \! \left(1-k \left[1+\frac{\left(\frac{q}{q -1}\right)^{q}\times \left[\omega  \hslash  \beta  q -\omega  \hslash  \beta -1\right] \times\left(\frac{1-\left(q -1\right) \omega  \hslash  \beta}{2 \left(q -1\right) \omega  \hslash  \beta}\right)^{-q}}{2 \omega  \hslash  \beta  \left(2 q -1\right)}\right]\right)}{q -1}
\end{equation}
\subsubsection{Temperature Behavioral Comparison}
Considering the structure of the obtained graphs, it is better to study the comparison of temperature-dependent behavior in two different ranges.\\ In the first step, we will look at normal and high temperatures, which are placed in the same category in terms of behavioral patterns.
\begin{figure}[H]
 \begin{center}
 \subfigure[]{
 \includegraphics[height=5.5cm,width=7cm]{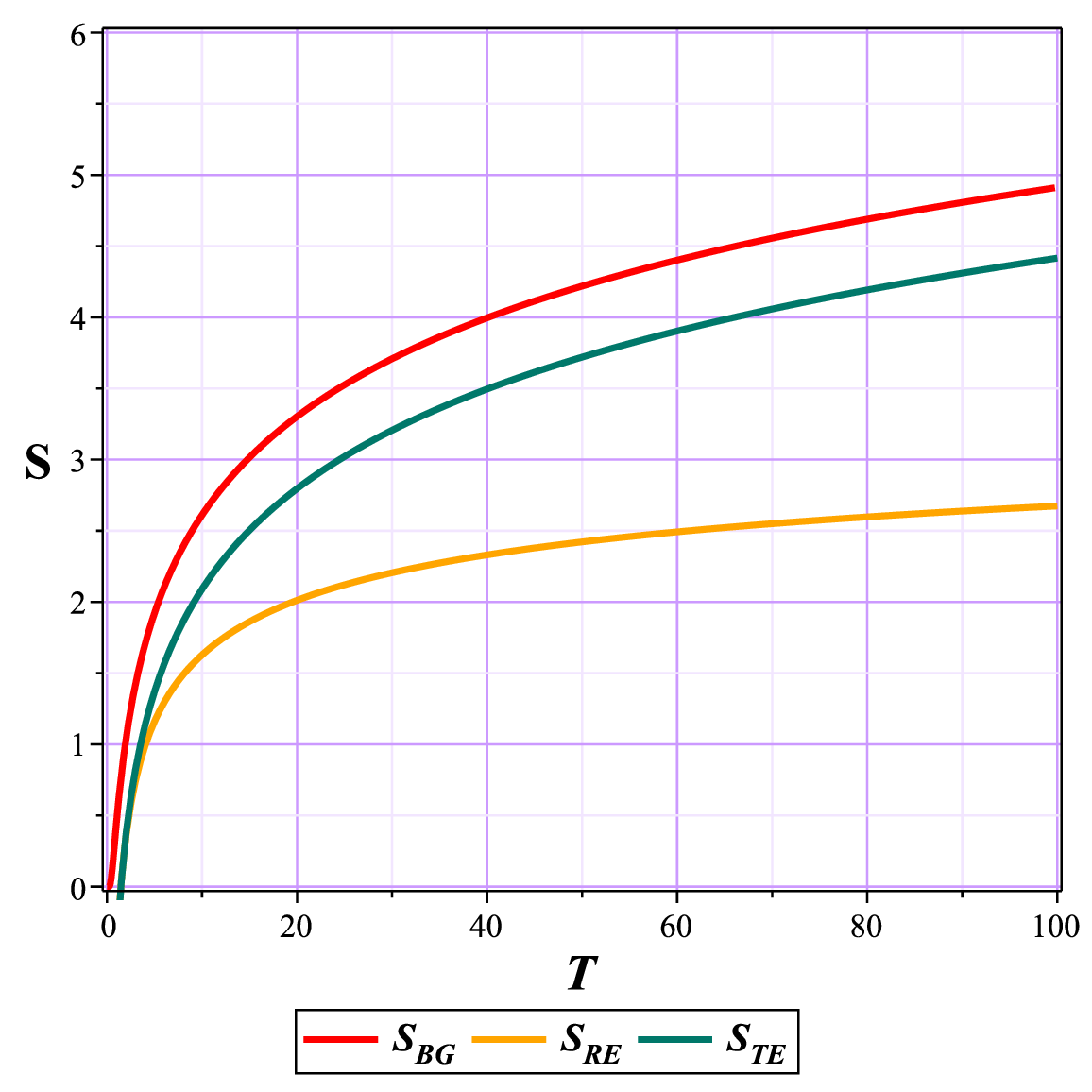}
 \label{2a}}
 \subfigure[]{
 \includegraphics[height=5.5cm,width=7cm]{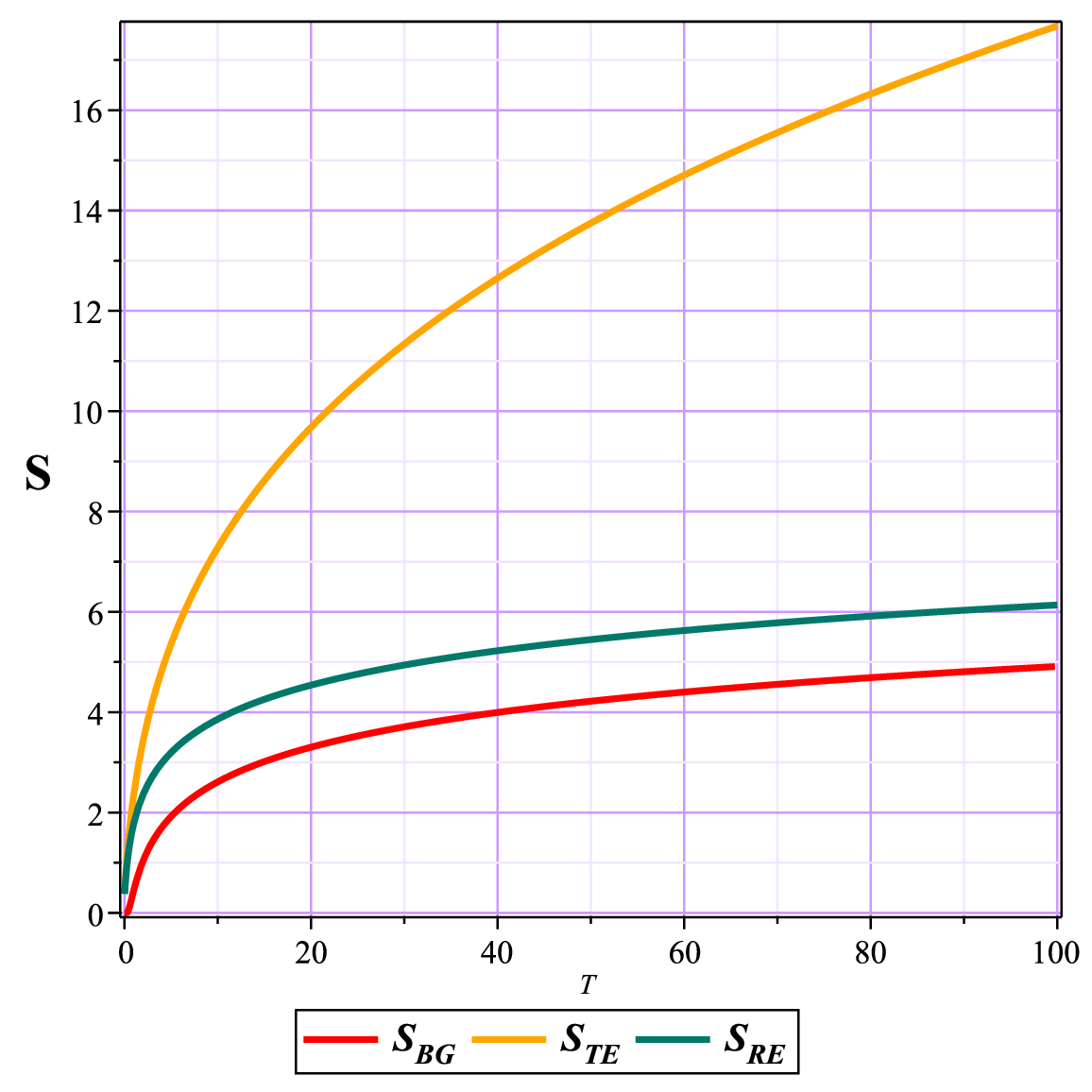}
 \label{2b}}
 \subfigure[]{
 \includegraphics[height=5.5cm,width=7cm]{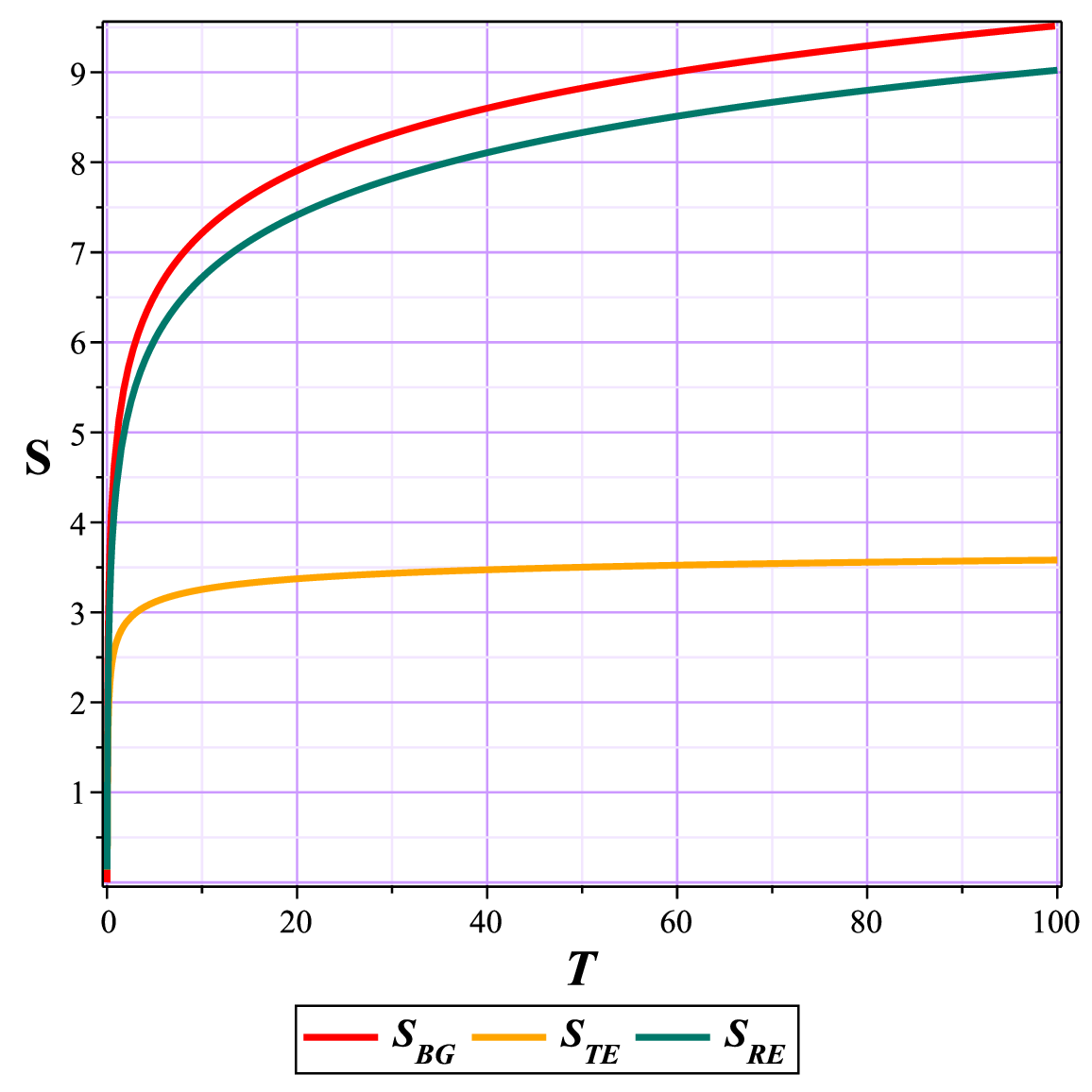}
 \label{2c}}
 \subfigure[]{
 \includegraphics[height=5.5cm,width=7cm]{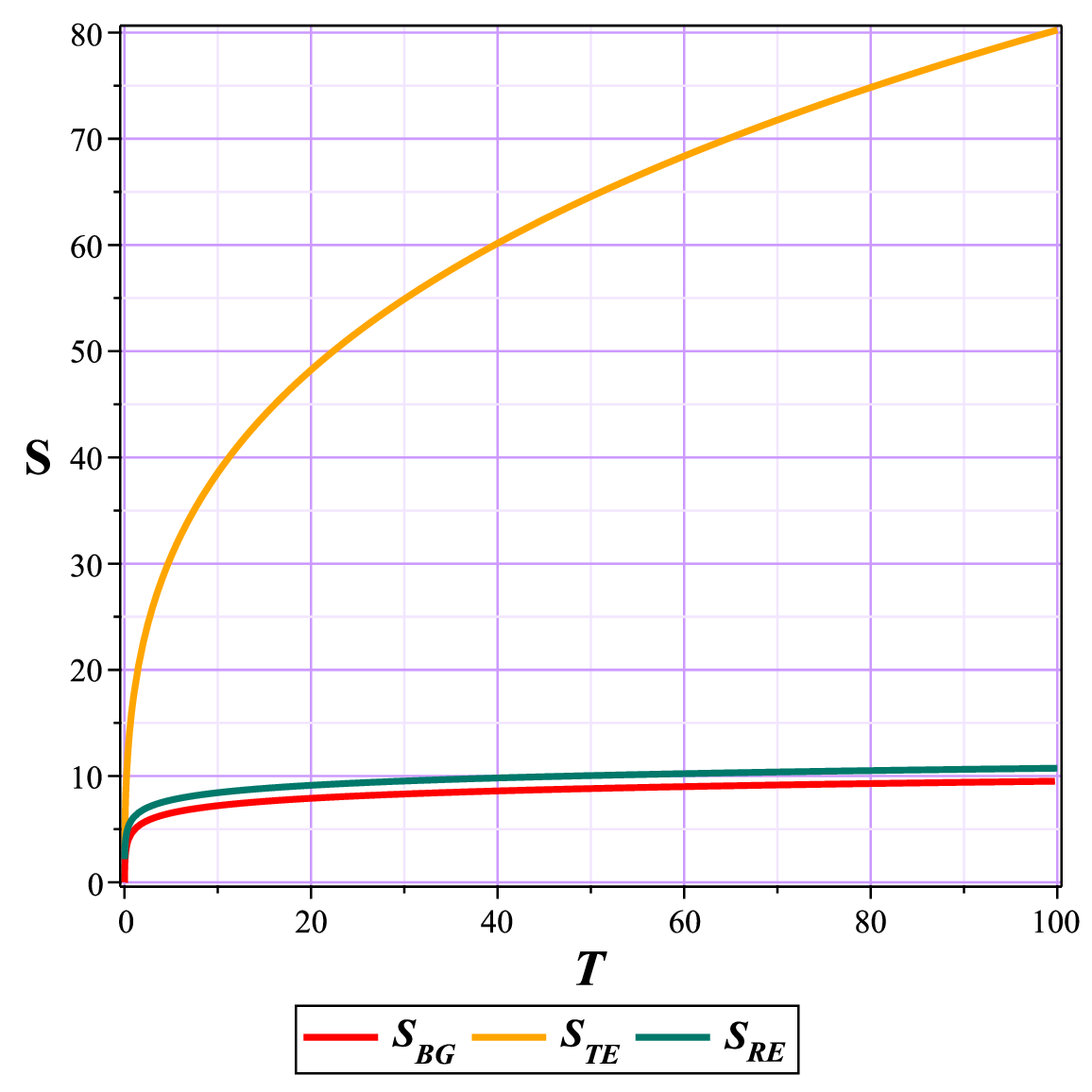}
 \label{2d}}
   \caption{\small{Temperature dependence of entropy in different form, at normal Temperature with  $\hbar=k=\omega=1$ , $q=1.25$ , $q=0.7$ in Fig (2a) and Fig (2b) respectively, and $\hbar=k=1$ and $\omega=0.01$ with $q=1.25$ and $q=0.7$ in Fig(2c), Fig(2d) .  }}
 \label{2}
\end{center}
\end{figure}
As shown in Fig. (\ref{2}), the deviation of the R\'enyi entropy from the Boltzmann entropy is markedly smaller than that of the Tsallis entropy, and this gap grows as the frequency \(\omega\) decreases. Thus, even with good accuracy (Fig. (\ref{2c}) and Fig. (\ref{2d})), the R\'enyi and Boltzmann results (even far from the standard limit state $q$ towards 1) can be considered to be in agreement to some extent and, depending on the needs of the study, one can be replaced by the other.
\subsubsection{Behavioral comparison at very low temperature}
At the beginning of this section, it is necessary to emphasize that, as stated in Section 3, we have used the precondition of high temperatures to derive our equations. Therefore, in this section, we do not intend to derive new relations by violating the high-temperature assumption. Instead, we examine the behavior of the obtained functions in the very low-temperature regime to investigate the extent to which the mathematical model, derived from the high-temperature approximation, remains consistent with standard results and reliable in regions where the initial assumption is not satisfied.
\begin{figure}[H]
 \begin{center}
 \subfigure[]{
 \includegraphics[height=5.5cm,width=7cm]{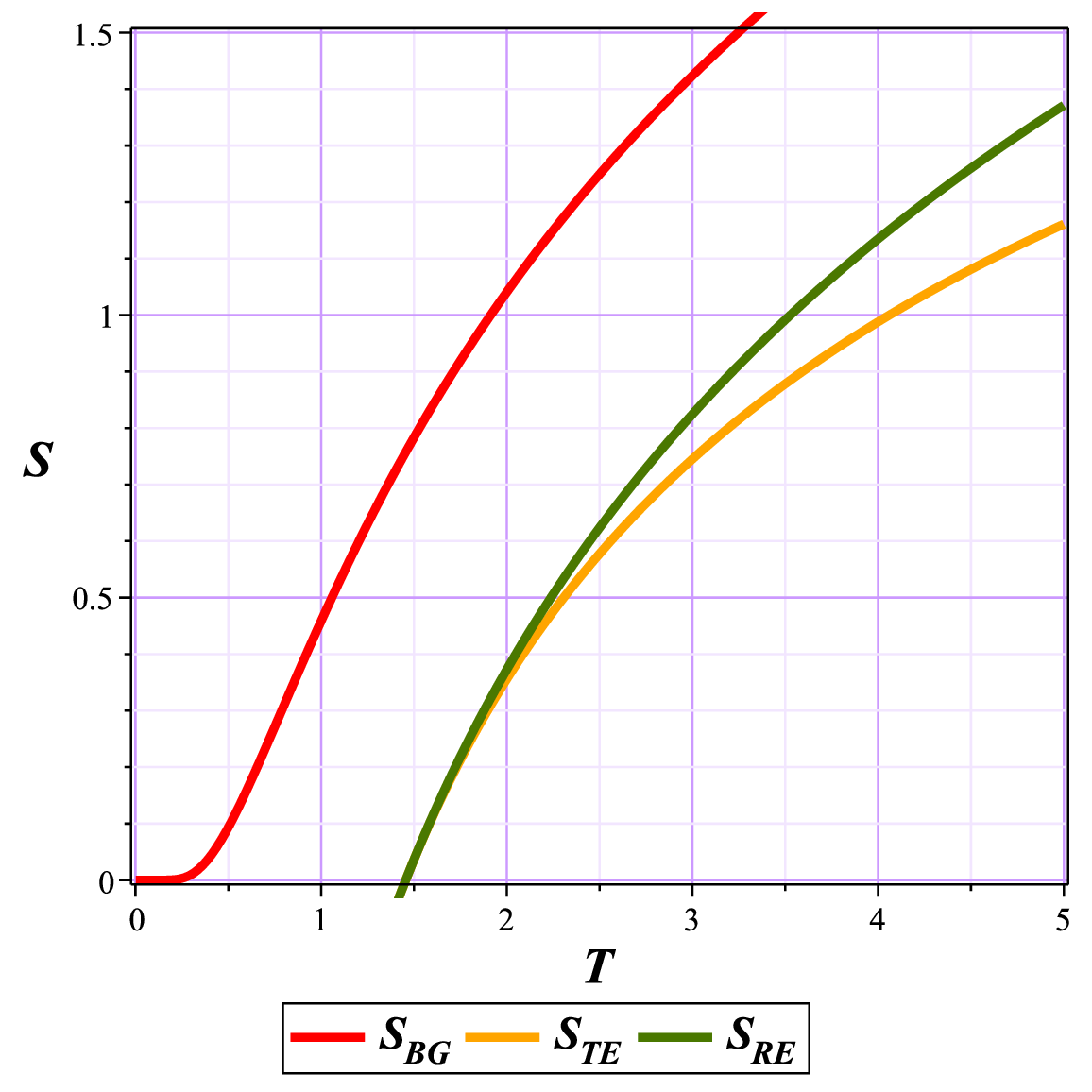}
 \label{3a}}
 \subfigure[]{
 \includegraphics[height=5.5cm,width=7cm]{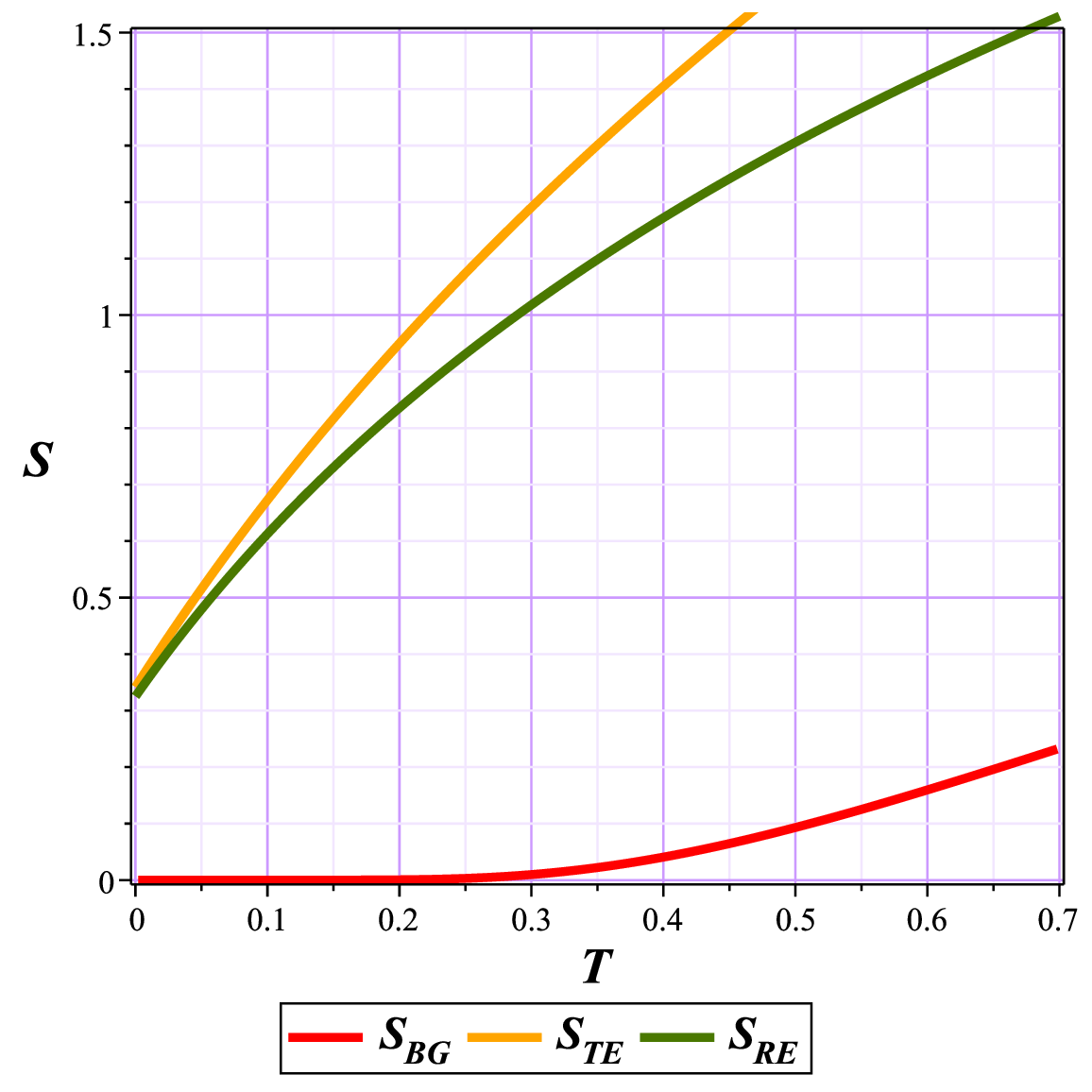}
 \label{3b}}
 \subfigure[]{
 \includegraphics[height=5.5cm,width=7cm]{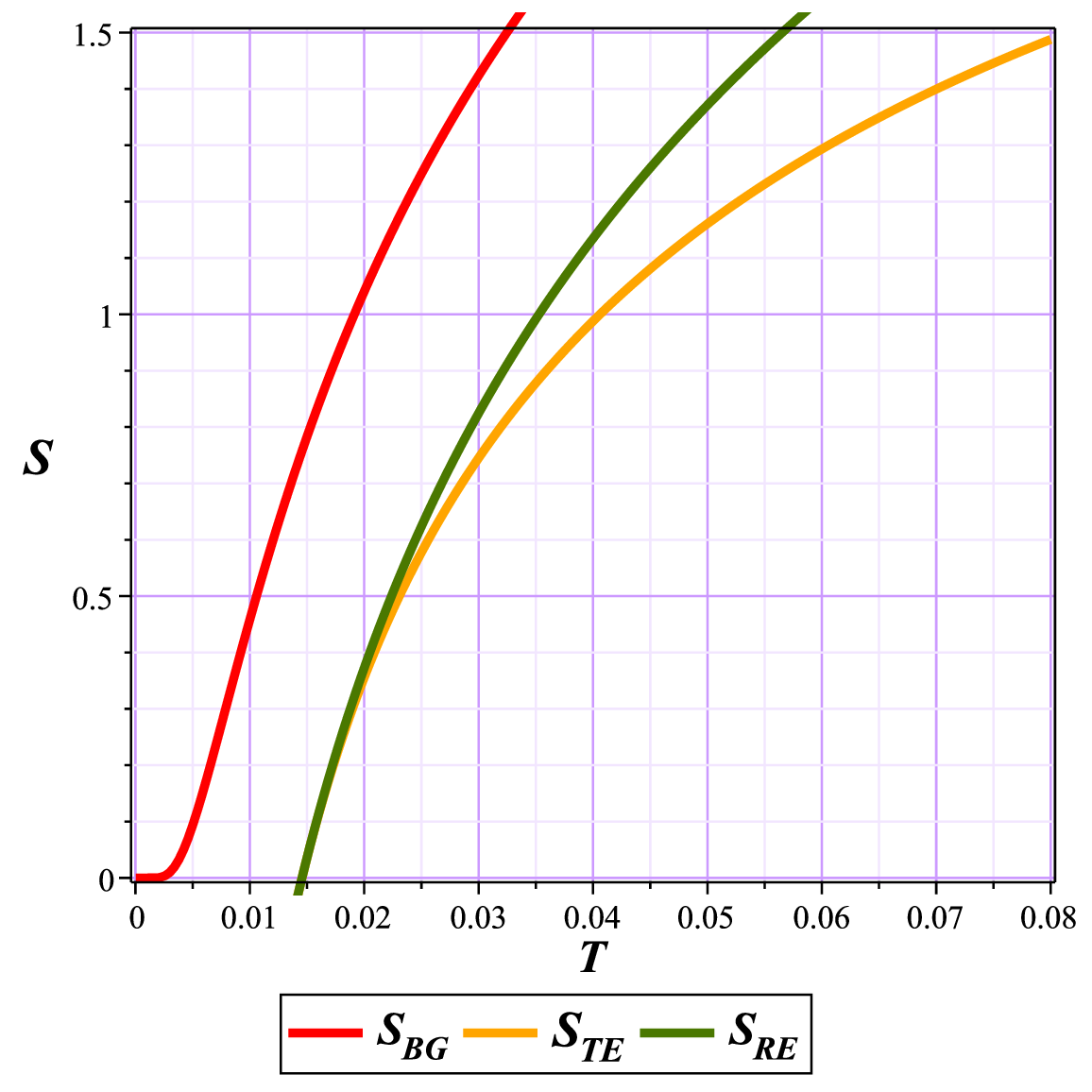}
 \label{3c}}
 \subfigure[]{
 \includegraphics[height=5.5cm,width=7cm]{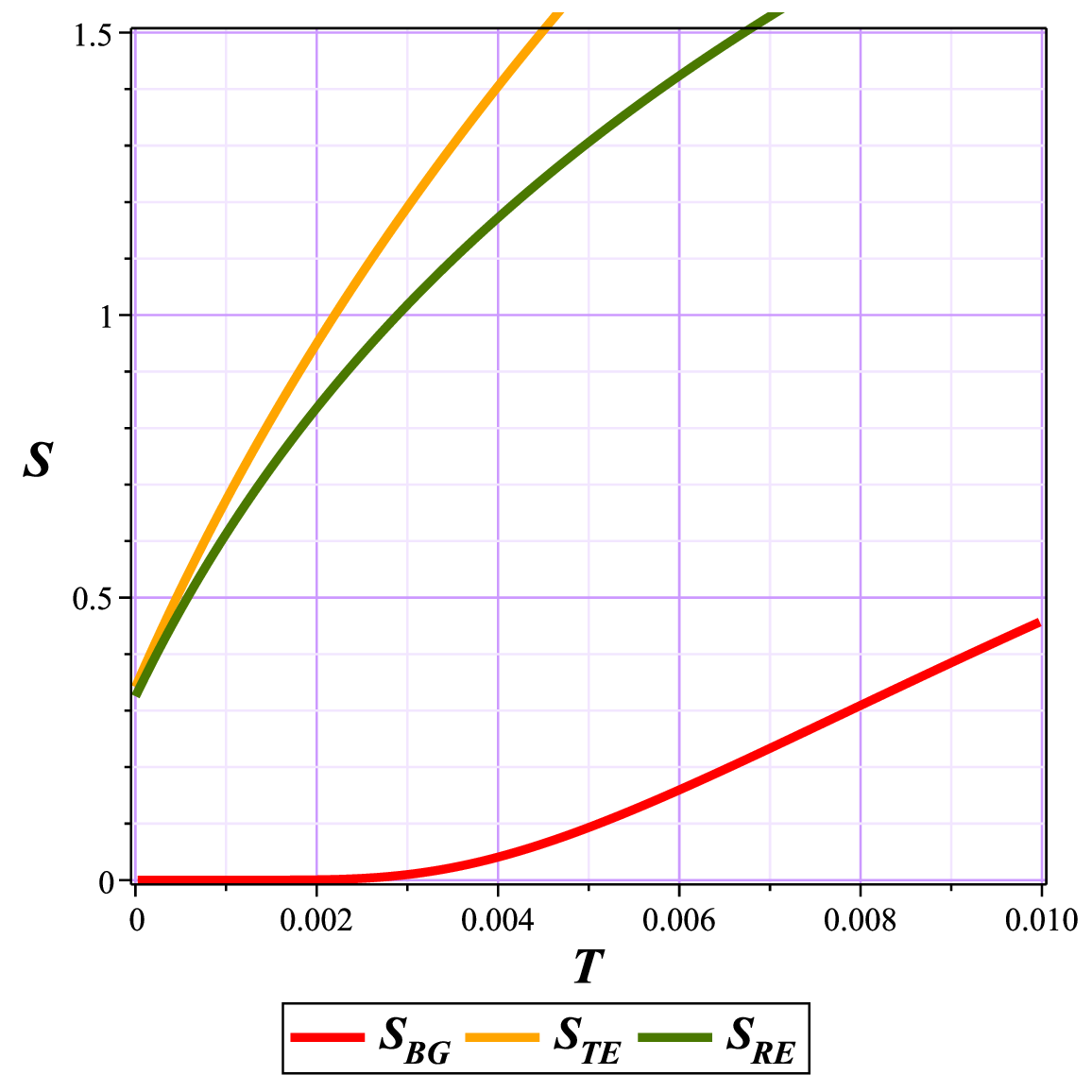}
 \label{3d}}
   \caption{\small{Temperature dependence of entropy in different form, at very low Temperature with  $\hbar=k=\omega=1$ , $q=1.25$ , $q=0.7$ in Fig (3a) and Fig (3b) respectively, and $\hbar=k=1$ and $\omega=0.01$ with $q=1.25$ and $q=0.7$ in Fig(3c), Fig(3d) . }}
 \label{3}
\end{center}
\end{figure}
As mentioned, although the high temperature approximation used is not completely valid for extrapolation to very low temperatures, Fig. (\ref{3}) shows that if we analytically extend these approximations to low temperatures, the Tsallis and R\'enyi entropies approach each other and deviate from the Boltzmann curve. This divergence may in some way indicate a discrepancy between the classical Boltzmann criterion and its generalized equivalents.\\
By definition, entropy quantifies a system’s disorder or randomness. Zero entropy denotes maximal order, when only a single, nondegenerate quantum microstate is accessible, a perfectly crystalline structure. Yet the Third Law of Thermodynamics dictates that absolute zero temperature cannot be reached: one may asymptotically approach it, but never attain it. Correspondingly, Boltzmann entropy asymptotically tends to zero temperature without ever completely vanishing, as reflected in all four panels of Fig. (\ref{3}).\\
Based on the approximate expressions, Fig. (\ref{3b}) and Fig. (\ref{3d}) indicate a finite positive value at $T=0$, which would violate the third law if taken literally. However, since our formulas were derived under the high-temperature condition $\beta\hbar\omega\ll1$ and they are not expected to be accurate near absolute zero, this apparent violation may be an artifact of extrapolating the approximate partition function beyond its domain of validity.\\  
Nevertheless, within their regime of validity (high temperature, small $|1-q|$), the generalized entropies provide a flexible phenomenological description. A fully consistent extension that satisfies the third law for all parameter values would require a more careful treatment beyond the present approximations.
\section{Statistical Comparison and Derivation of the Relation Between $q$ and $\theta$}
For the two-dimensional harmonic oscillator in a noncommutative space as we seen The energy eigenvalues are given by Eq. (\ref{15}). In the symmetric case where $n_x = n_y = n$, the spectrum simplifies to:
\begin{equation}
E_n = 2 \hbar \Omega \left(n + \frac{1}{2} \right),
\label{27}
\end{equation}
and for the partition function in this symmetric case we have:
\begin{equation}
Z_{\theta} = \sum_{n=0}^{\infty} e^{-\beta E_n} = \frac{e^{-\beta \hbar \Omega}}{1 - e^{-2\beta \hbar \Omega}}.
\label{28}
\end{equation}
Now the entropy can be calculated by using the canonical definition from Boltzmann-Gibbs statistics:
\begin{equation}
S_{\theta} = k \left[ \frac{2\beta \hbar \Omega}{e^{2\beta \hbar \Omega} - 1} - \ln \left(1 - e^{-2\beta \hbar \Omega}\right) \right].
\label{29}
\end{equation}
By comparing this equation with Eq. (\ref{24}), it becomes evident that despite the presence of spatial non commutativity, the thermodynamic quantities retain their structural pattern analogous to the commutative case under the condition \( n_x = n_y \). The key distinction lies in the replacement of the original frequency \( \omega \) with an effective frequency \( \Omega \), wherein the noncommutative parameter \( \theta \) is implicitly encoded in the \( \Omega \). Consequently, the influence of non commutativity manifests through this renormalized frequency, thereby affecting the system's thermodynamic behavior. From this perspective, the effect of noncommutativity on the entropy can be interpreted as an effective modification of the oscillator frequency. Therefore, the deviations observed in Fig. (\ref{1}) for different values of the generalized parameter $q$ provide a useful qualitative picture of how a deformation parameter can alter the entropy profile relative to the Boltzmann–Gibbs result. Although the detailed correspondence between $q$ and the noncommutative parameter $\theta$ will be derived later, both parameters ultimately manifest themselves through modifications of the entropy curves.
This synthesis naturally prompts whether such a formulation can be systematically employed to investigate the impact of non commutativity on thermodynamic properties, potentially establishing a novel and significant connection in the theoretical framework.\\
For this aim, we begin with the approximate form of the Tsallis partition function derived earlier Eq. (\ref{19}) which for it in the limit $\beta \hbar \omega (q - 1) \ll 1$ we have:
\begin{equation}
Z_q \approx \frac{1}{2 \beta \hbar \omega q} - \frac{1}{2}
\label{30}
\end{equation}
Also, the canonical partition function for the oscillator in noncommutative space, in the symmetric case $n_x = n_y = n$ and at high temperature, simplifies to:
\begin{equation}
Z_\theta \approx \frac{1}{2 \beta \hbar \Omega} - \frac{1}{2},
\label{31}
\end{equation}
where $\Omega$ is the effective frequency incorporating noncommutativity:
\begin{equation}
\Omega = \omega \sqrt{1 + \frac{m^2 \omega^2 \theta^2}{4 \hbar^2}}.
\label{32}
\end{equation}
Comparing the two above equations leads to :
\begin{equation}
\frac{1}{2 \beta \hbar \omega q} \approx \frac{1}{2 \beta \hbar \Omega},
\label{33}
\end{equation}
which immediately implies the relation between $q$ and $\Omega$:
\begin{equation}
\Omega \approx \omega q.
\label{34}
\end{equation}
Then, using Eq. (\ref{32}), one can derive the explicit relation between $q$ and $\theta$:
\begin{equation}
q = \sqrt{1 + \frac{m^2 \omega^2 \theta^2}{4 \hbar^2}}
\label{35}
\end{equation}
We emphasize that this relation is valid in the high-temperature limit ($\beta\hbar\omega \ll 1$) and for small noncommutativity ($m\omega\theta/\hbar \ll 1$). Outside this regime, higher-order corrections become important.\\As previously demonstrated, the comparison of Eq. (\ref{24}) and Eq. (\ref{29}) shows that, even in the noncommutative regime, the entropy’s temperature dependence virtually replicates the simple harmonic‐oscillator profile. A close inspection of Equation  Eq. (\ref{29}) further reveals that as the noncommutativity parameter $\theta$ decreases, this correspondence becomes increasingly  exact. 
Before discussing the corresponding R\'enyi behavior, it is useful to recall that the relation established above connects the noncommutative entropy to the Tsallis framework through the parameter mapping derived in Eq. (\ref{33})--Eq. (\ref{35}). Since the R\'enyi entropy is directly related to the Tsallis entropy through  Eq. (\ref{(25)}), any qualitative correspondence inferred from the Tsallis description can also be translated into the R\'enyi framework. This provides the motivation for comparing the noncommutative entropy with the R\'enyi entropy in the following discussion.
Hence, without additional plotting and by reference to Fig. (\ref{2}), one can conclude that the R\'enyi entropy will exhibit smaller deviations from the corresponding noncommutative entropy and will lie closer to it. To provide a more intuitive physical interpretation of the derived correspondence, one may directly refer to Fig. (\ref{1}). According to the mapping obtained in Eq. (\ref{35}), a value of $q>1$ corresponds to a nonzero noncommutative parameter $\theta$. Therefore, comparing the $S_{BG}$ curve with the $S_{q=1.25}$ curve in Fig. (\ref{1b}) provides a qualitative illustration of how noncommutativity may affect the entropy behavior. In this picture, the noncommutative deformation is encoded through the effective frequency $\Omega$, leading to a modification of the entropy relative to the commutative Boltzmann--Gibbs result. For the representative case corresponding to $q>1$, the entropy curve lies below the $S_{BG}$ curve, indicating a reduction of entropy at a fixed temperature. Since Eq. (\ref{35}) was derived in the high-temperature and weak-noncommutativity regime; this comparison should be interpreted as a qualitative physical illustration rather than an exact quantitative correspondence for arbitrary values of $q$. Therefore, Fig.~1 may be viewed as providing an intuitive visualization of the thermodynamic consequences of the $q$--$\theta$ correspondence established in Eq. (\ref{35}).

\begin{figure}[H]
\centering
\subfigure[]{
 \includegraphics[height=5.5cm,width=7cm]{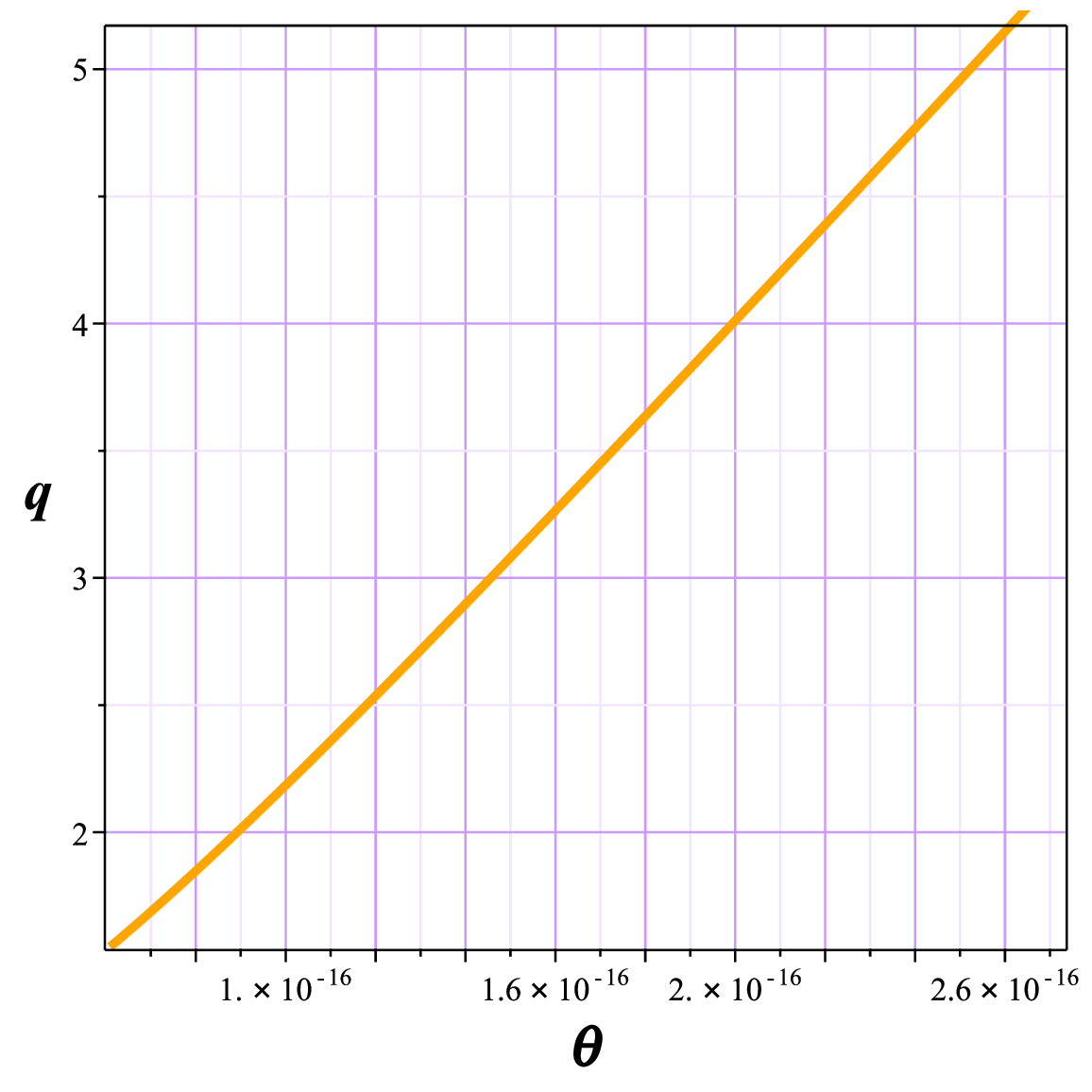}
 \label{4a}}
 \subfigure[]{
 \includegraphics[height=5.5cm,width=7cm]{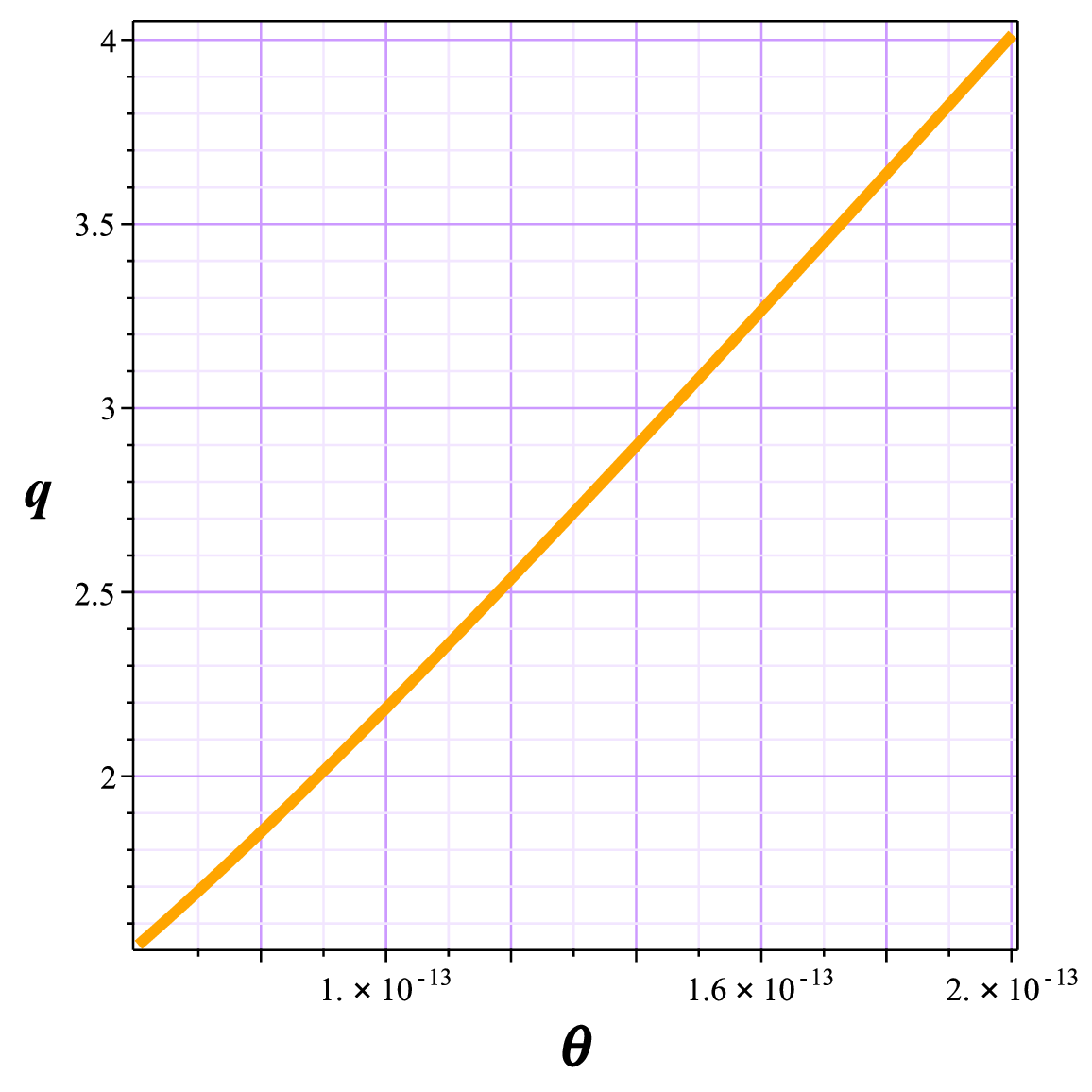}
 \label{4b}}
 \subfigure[]{
 \includegraphics[height=5.5cm,width=7cm]{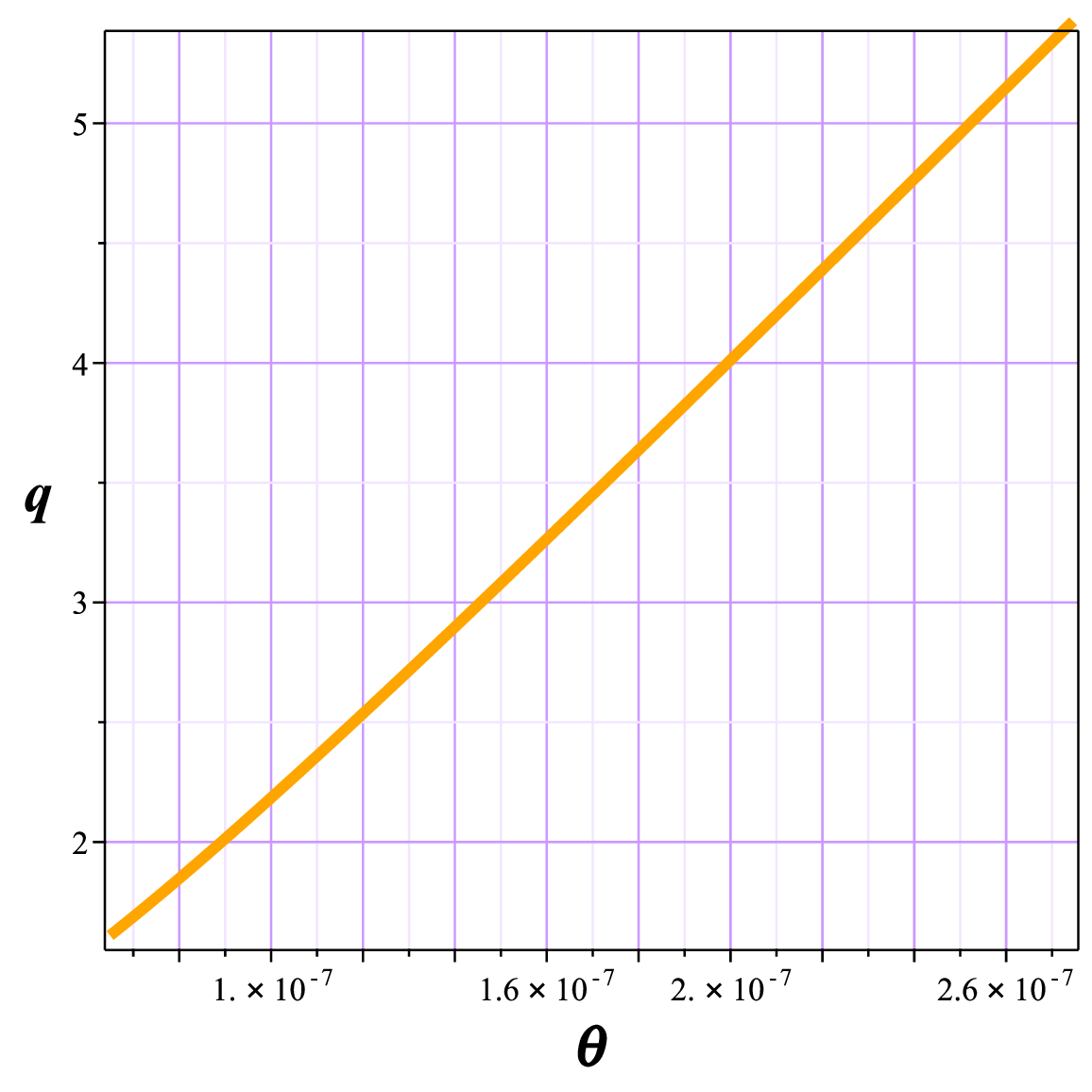}
 \label{4c}}
\caption{The analytical relation between the nonextensive parameter $q$ and the noncommutative parameter $\theta$.  Fig (4a) for $ \hbar=6.626*10^{-34} ,m \omega=1$ ,Fig (4b) for $ \hbar=6.626*10^{-34} ,m \omega=10^{-3} $, Fig (4c) for $ \hbar=6.626*10^{-34} ,m \omega=10^{-9} $} 
\label{4}
\end{figure}
Unlike Fig. (\ref{1})--Fig. (\ref{3}), where Planck's constant was set to unity since it only affected the graphical scale, Fig. (\ref{4}) is plotted using SI units for a more precise representation.\\
Examination of Eq.~(\ref{35}) reveals a smooth monotonic relationship between the Tsallis entropic index $q$ and the noncommutative parameter $\theta$, as illustrated in Fig.~(\ref{4}). Although the exact dependence is nonlinear, over sufficiently restricted intervals of $\theta$ the variation may appear nearly linear. It should be emphasized that this correspondence holds in the high-temperature regime $\beta \hbar \omega \ll 1$ and for small non-commutativity $m\omega\theta/\hbar \ll 1$. Beyond this regime, corrections of order $(\beta \hbar \omega)^2$ and $(m\omega\theta/\hbar)^4$ become significant, and the exact partition functions must be used for a more precise mapping.
The purpose of Fig. (\ref{4}) is to visualize the correspondence between the Tsallis entropic parameter $q$ and the noncommutative parameter $\theta$ implied by Eq.~(\ref{35}). Although the exact relation is nonlinear, the figure illustrates that $q$ increases monotonically with $\theta$, thereby establishing a unique correspondence between the physically relevant values of $\theta$ and the parameter $q$ within the validity of the approximation. The different panels correspond to different values of the combination $m\omega$, showing how the sensitivity of $q$ to variations in $\theta$ depends on the physical scale of the oscillator. Therefore, the significance of Fig. (\ref{4}) is not to demonstrate strict linearity, but rather to provide a graphical interpretation of the parameter correspondence that underlies the connection between noncommutative thermodynamics and generalized entropic formalisms.
This $q$--$\theta$ correspondence carries profound implications, for it forges a powerful bridge between two ostensibly distinct physical notions. The nonextensive deformation introduced by $q$ in Tsallis statistics effectively models the geometric distortion induced by $\theta$, suggesting that the Tsallis index can serve as an effective parameter for spatial noncommutativity in this regime.\\
Furthermore, as shown in Fig. (\ref{2d}), selecting an appropriate $q$ and operating in the low–frequency regime brings the R\'enyi entropy curve very close to its noncommutative counterpart, so that in practice one description can be substituted for the other with good accuracy.
In fact, by interpreting $q$ from a R\'enyi rather than a Tsallis perspective, R\'enyi entropy-based analyses can be systematically translated into the language of noncommutativeness, and vice versa, opening new avenues of study.
An intriguing observation concerns the two-parameter Sharma–Mittal entropy, defined by \(r\) and \(q\) as \cite{11}:
\[
S_{r,q} \;=\; \frac{1}{r - 1}\,\biggl[\Bigl(\sum_i p_i^q\Bigr)^{\frac{r - 1}{q - 1}} \;-\; 1\biggr].
\]

In the limit \(r \to 1\), this measure reduces to the Rényi entropy, suggesting that—at least in this boundary case—one might likewise encode the noncommutativity parameter \(\theta\) within the Sharma–Mittal framework. Such an extension would further generalize the deep connection between nonextensive statistics and spatial noncommutativity.

\section{ Important Discussion }
Before ending, it is better to point out a fundamental point that may confuse the reader's mind, which is that we used the symmetric precondition \(n_x = n_y\) for our study in noncommutative space and to obtain the relations in Section 5. Indeed, in the symmetric sector \(n_x = n_y\), the partition function of the non-commutative harmonic oscillator reduces formally to that of a commutative oscillator with a renormalized frequency \(\Omega(\theta)\). But this simplification was adopted to obtain analytic tractability and to isolate the parametric role of \(\theta\) in a more simple setting.\\
However, the fact that non-commutativity enters as a renormalization of \(\omega\) does not diminish the physical significance of the connection we found. Rather, it reveals a key interpretative insight: the non-commutative parameter \(\theta\) behaves as a geometric deformation that can be absorbed into an effective dynamical scale (\(\Omega\)), much like the entropic index \(q\) acts as a deformation parameter in generalized statistics. The mapping \(q \approx \sqrt{1 + \frac{m^2 \omega^2 \theta^2}{4\hbar^2}}\) thus establishes a 'dictionary-like' correspondence between a geometric deformation (spacetime non-commutativity) and a statistical deformation (nonextensivity).\\
In this sense, the “deep link” we propose is not about discovering entirely new thermodynamic structures in the symmetric sector, but about demonstrating that two ostensibly independent deformations—one geometric, one statistical—can produce thermodynamically equivalent effects when viewed through the lens of effective parameters. This equivalence suggests that in certain regimes, non-commutative effects might be 'mimicked' or 'modeled' by nonextensive statistics, and vice versa.\\
Moreover, the symmetric case serves as a controlled starting point. Future work may extend this analysis to the full non-symmetric spectrum $n_x \neq n_y$, where degeneracy lifting introduces additional thermodynamic features beyond simple frequency renormalization. Nevertheless, in this more general setting, the parametric correspondence between $\theta$ and $q$ is expected to persist at leading order, while higher-order corrections would capture the finer aspects of noncommutative thermodynamics.\\
Thus, while we acknowledge that our present thermodynamic analysis is confined to a sector where non-commutativity manifests mainly through \(\Omega\), this very simplicity allows us to isolate and highlight a clean, interpretable correspondence between \(\theta\) and \(q\). This correspondence lays a conceptual foundation for exploring how geometric and statistical deformations interplay in more complex, physically richer regimes.\\
\section{ Conclusion }
Although the Boltzmann–Gibbs entropy underpins classical statistical mechanics, its applicability diminishes in systems characterized by long-range interactions, memory effects, or fractal phase spaces. Over time, these limitations prompted the development of alternative generalized entropy measures—such as Tsallis, R\'enyi, Shannon, and Sharma–Mittal entropies—introduced progressively to capture nonstandard statistical behavior. We now appreciate that analyzing both canonical classical structures and advanced modern models using these generalized entropies yields profound insights. In quantum mechanics, one widespread method for dissecting quantum systems involves mapping their Hamiltonians onto the Hamiltonian of a harmonic oscillator, whose well-known energy spectrum serves as a golden key linking the oscillator’s dynamics to statistical mechanics via entropy.\\
Motivated by this approach, we considered the simple harmonic oscillator in two contrasting geometric frameworks: the usual commutative (classical) space and a noncommutative space. Employing the energy-spectrum correspondence, we conducted a thermal analysis of the oscillator’s entropy in both the Tsallis and R\'enyi formalisms. Our comparative investigation revealed that, at moderate to high temperatures, R\'enyi entropy converges closely to the Boltzmann–Gibbs result. Although examining the behavior of Eq. (\ref{19}) and Eq. (\ref{21}) at low temperatures goes beyond the initial approximations used to derive these equations, it can be mathematically observed that, near absolute zero, the Tsallis and R\'enyi entropies approach each other while deviating from the Boltzmann--Gibbs prediction. This apparent divergence arises because the high-temperature approximation underlying these expressions is no longer valid in this regime, and should therefore be interpreted as a mathematical artifact rather than a physically reliable result.\\
The primary objective of this work is to explore how spacetime noncommutativity influences these generalized entropies, thereby opening a new research direction. In particular, we probe whether a fundamental relationship exists between the noncommutativity parameter $\theta$ and the entropy index $q$—whether in the Tsallis framework or its R\'enyi generalization—and how such a connection might unveil deep interplay between spacetime geometry and statistical mechanics.\\
Our analysis indicates a clear monotonic correspondence between the Tsallis index $q$ and the noncommutativity parameter $\theta$. This $q$--$\theta$ correspondence has profound implications: it establishes a robust bridge between geometric (noncommutative) and statistical (nonextensive) deformations without implying strict linearity.\\ 
The nonextensive deformation encoded by $q$ in Tsallis and R\'enyi statistics effectively models the geometric distortion induced by $\theta$, providing an integrated interpretive framework in which the Tsallis index acts as an effective parameter representing spatial noncommutativity in the high-temperature and weak noncommutativity regime. Furthermore, given the analogous entropy behavior of the commutative and noncommutative oscillators (subject to the appropriate mapping of $\omega$ to $\Omega$) and the convergence of Boltzmann–Gibbs and R\'enyi entropies, one can systematically translate R\'enyi-based analyses into the language of noncommutative geometry—and vice versa—thus opening exciting new avenues for exploration. Additionally, the limiting-case correspondence between R\'enyi and other generalized entropies, such as Sharma–Mittal, suggests promising links between multiparameter entropy measures and noncommutativity.\\
Finally, it is worth comparing the present results with the thermodynamic behavior reported in previous studies based on standard Boltzmann–Gibbs statistics. In \cite{Jahan2012}, the partition function of the two-dimensional noncommutative harmonic oscillator was derived using a path integral approach at finite temperature, showing that noncommutativity modifies the energy spectrum and consequently the thermodynamic quantities through an effective frequency. Similarly, \cite{Lin2018} investigated the entanglement entropy of harmonic oscillators in noncommutative phase space, demonstrating that noncommutativity can induce entanglement-like effects. Our results, obtained within the framework of generalized entropies, are consistent with these findings in the sense that the noncommutative parameter $\theta$ manifests as a renormalization of the oscillator frequency $\Omega$. However, by introducing the Tsallis and R\'enyi entropies, we have shown that the statistical deformation parameter $q$ can be directly related to the geometric deformation parameter $\theta$, providing a complementary perspective that bridges noncommutative geometry and nonextensive statistical mechanics.\\
Collectively, these insights deepen our understanding of statistical mechanics in noncommutative geometries and furnish a robust framework for modeling complex systems endowed with intrinsic geometric or statistical deformations.\\\\
\begin{center}
\textbf{Acknowledgements}
\end{center}
We sincerely thank the referee for their careful comments and suggestions, which helped to improve the quality of the article.\\
This research is supported by the research grant of the University of Mazandaran (number: 33/66365).

\end{document}